%% file: main-arxiv.tex
\documentclass[conference]{IEEEtran}
\IEEEoverridecommandlockouts
\usepackage{cite}
\usepackage{amsmath,amssymb,amsfonts}
\usepackage{graphicx}
\usepackage{textcomp}
\usepackage{xcolor}
\usepackage{courier}

\usepackage[]{graphicx}
\usepackage{pseudocode}
\usepackage[figuresright]{rotating}
\usepackage{subcaption}
\usepackage{listings}

\usepackage{booktabs}
\usepackage{tabularx}
\usepackage{caption}
\usepackage{multicol}
\usepackage{multirow}
\usepackage{xspace}
\usepackage[export]{adjustbox}

\usepackage{colortbl}

\usepackage{pifont}

\usepackage{breakurl}
\usepackage[hyphens]{url}

\usepackage[]{algorithm2e}

\usepackage{amsmath}
\usepackage{mathtools}
\usepackage[
    n,
    operators,
    advantage,
    sets,
    adversary,
    landau,
    probability,
    notions,
    logic,
    ff,
    mm,
    primitives,
    events,
    complexity,
    asymptotics,
    keys]{cryptocode}

\usepackage{fancyhdr}
\usepackage{hhline,balance}
\usepackage{datetime}
\usepackage{stfloats}
\usepackage{blkarray}

\newcommand{\squishlist}{
 \begin{list}{$\bullet$}
  { \setlength{\itemsep}{0pt}
     \setlength{\parsep}{3pt}
     \setlength{\topsep}{3pt}
     \setlength{\partopsep}{0pt}
     \setlength{\leftmargin}{1.5em}
     \setlength{\labelwidth}{1em}
     \setlength{\labelsep}{0.5em} } }

\newcommand{\squishlisttwo}{
 \begin{list}{$\bullet$}
  { \setlength{\itemsep}{0pt}
     \setlength{\parsep}{0pt}
    \setlength{\topsep}{0pt}
    \setlength{\partopsep}{0pt}
    \setlength{\leftmargin}{2em}
    \setlength{\labelwidth}{1.5em}
    \setlength{\labelsep}{0.5em} } }

\newcommand{\squishend}{
  \end{list}  }

\newcommand{\our}{\textsf{Sash}}

\DeclareCaptionFont{white}{\color{black}} 
\DeclareCaptionFormat{listing}{\colorbox{white}{\parbox{0.9\textwidth}{#1#2#3}}} 
\graphicspath{{../fig/}}

\PassOptionsToPackage{hyphens}{url}

\makeatletter
\def\ps@IEEEtitlepagestyle{%
  \def\@oddfoot{\mycopyrightnotice}%
  \def\@evenfoot{}%
}
\def\mycopyrightnotice{%
  {\footnotesize \textit{Proceedings of the 2nd IEEE International Conference on Blockchain, Atlanta, USA, 2019 \hfill}}
  \gdef\mycopyrightnotice{}
}

\begin{document}

\title{Towards Secure and Decentralized \\ Sharing of IoT Data}

\author{\IEEEauthorblockN{Hien Thi Thu Truong, Miguel Almeida, Ghassan Karame, Claudio Soriente}
\IEEEauthorblockA{\textit{NEC Laboratories Europe}, Germany \\
hien.truong@neclab.eu, miguel.almeida@neclab.eu, ghassan.karame@neclab.eu, claudio.soriente@emea.nec.com}
}

\maketitle

\begin{abstract}
\input{abstract}
\end{abstract}


\input{intro}
\input{background}

\input{system}

\input{sharing}
\input{evaluation}
\input{related}
\input{conclusion}

\input{ack}

\bibliographystyle{abbrv}
\bibliography{cited}

\end{document}

%% file: abstract.tex
The Internet of Things (IoT) bears unprecedented security and scalability challenges due to the magnitude of data produced and exchanged by IoT devices and platforms. Some of those challenges are currently being addressed by coupling IoT applications with blockchains. However, current blockchain-backed IoT systems simply use the blockchain to store access control policies, thereby underutilizing the power of blockchain technology. In this paper, we propose a new framework named \our{} that couples IoT platforms with blockchain that provides a number of advantages compared to state of the art. In \our{}, the blockchain is used to store access control policies and take access control decisions. Therefore, both changes to policies and access requests are correctly enforced and publicly auditable. Further, we devise a ``data marketplace'' by leveraging the ability of blockchains to handle financial transaction and  providing ``by design'' remuneration to data producers. Finally, we exploit a special flavor of identity-based encryption to cater for cryptography-enforced access control while minimizing the overhead to distribute decryption keys. We prototype \our{} by using the FIWARE open source IoT platform and the Hyperledger Fabric framework as the blockchain back-end. We also evaluate the performance of our prototype and show that it incurs tolerable overhead in realistic deployment settings.

%% file: intro.tex
\section{Introduction}

The Internet of Things (IoT) has been envisioned as a large distributed system of devices equipped with sensors and actuators. IoT devices are expected to create and exchange vast amounts of data, thereby bringing forth unprecedented challenges in terms of security and scalability. In such particular settings, available solutions for secure data sharing fall short owing to the magnitude of data produced, heterogeneity of devices, lack of trust among parties, and transparency on data handling.

In many distributed applications where trust and transparency are critical factors, the blockchain technology has shown to be a promising solution. It is not surprising, therefore, that both industry and research community are heavily discussing on how to efficiently combine IoT platforms with blockchains. For example, a number of industrial players are developing blockchains tailored to IoT use cases (e.g. IOTA\footnote{\url{https://www.iota.org/}}, IoTeX\footnote{\url{https://iotex.io/}}, Atonomi\footnote{\url{https://atonomi.io}}).
Similarly, a number research proposals suggest to solve the problem of secure data sharing in IoT platforms by directly connecting them to a blockchain platform~\cite{cryptoeprint:2018:209, Christidis16, Shrestha18, Laurent2018AnAC}. The majority of these proposals follow a hybrid approach where a storage system (e.g., a cloud provider) hosts the data itself and a companion blockchain offers services to ensure, e.g., trust distribution and integrity. For example,~\cite{Shafagh:2017:TBA:3140649.3140656} propose to store access control policies that are queried by the storage provider whenever it receives an access request. As such, the storage provider acts as a policy decision and enforcement point; and the blockchain ensures policy integrity and allows a public auditing of the changes made to a policy. Other proposals~\cite{Hu2018, Kaaniche2017} assume data to be encrypted before being stored at the cloud, and introduce an additional key-authority to regulate access to keys needed to decrypt data. In such a scenario, the damages of a malicious storage provider are mitigated as it only handles ciphertexts. Distribution of the key authority~\cite{Yakubov} further avoids single point of failure for handling decryption keys.

Nevertheless, existing proposals are not fully solving scalability for access control problems taking account the vast number of IoT stakeholders and potential sharing transactions among them. They propose to port access policies to the blockchain, which make trust distributed, but still they require owners to handle policy updates. This mechanism does not scale to the size of IoT systems. We argue that blockchain can enable this policy management to be more scalable by shifting policy update operations to the blockchain back-end via implementation of smart contracts. By doing so, we are taking full advantage of blockchain technology for IoT. On the one hand, blockchains rely on consensus engine to maintain the integrity of a ledger and to audit its changes---so it can be used to hold the ``true'' access control policy for a piece of data. On the other hand, blockchains are also an effective means to (1) ease the setup of communication channels between parties, (2) monetize transactions and information exchange, and (3) perform general computation by means of smart contracts. 

We therefore propose a novel blockchain-enhanced IoT data-sharing framework named \our{} that takes full advantage of the provisions offered by the blockchain. We follow the established paradigm of storing data offchain---given the amount of data in IoT applications, no other option seems viable. However, we shift more operations to the blockchain back-end thereby exploiting its natural resilience to malicious behavior of its members. We instantiate our proposed framework with  a number of tools borrowed from system security and applied cryptography, leading to solutions with different level of trust in the off-chain components. In our basic instantiation, we place the policy decision point (PDP) in the blockchain by leveraging smart contracts. A dedicated contract handles access control policies and evaluates access requests. Data owners may set a price to access their data and access decisions take into account whether the data consumer is willing to pay that price. As such, (1) policies are correctly and fairly evaluated, (2) granting access decisions can be audited, and (3) compensation to data producer is offered ``by design''. The cloud storage, therefore, merely acts as a policy enforcement point (PEP).

In our extended instantiation, we refine the trust asumption on the cloud storage and assume data is encrypted by data owners before being stored in the cloud. In particular, we use \emph{prefix encryption}~\cite{Boneh:2006:CSI:1272948.1272952} to distribute decryption keys. Prefix encryption is a special flavor of Hierarchical Identity-Based Encryption (HIBE)~\cite{Boneh:2005:HIB:2154598.2154634} and it is particularly suited for applications where data is arranged in a hierarchy. 
As such, prefix encryption allows fine-grained access control and minimizes the overhead required to issue/obtain decryption keys. The key distribution authority may be run by the owner of the sensors, by a trusted authority, or even distributed across several parties to avoid a single point of failure.

We prototype \our{} using FIWARE as the IoT platform and the Hyperledger Fabric as the blockchain back-end. The essential parts of \our{} are two new handlers \texttt{Blockchain Handler} and \texttt{IoT Domain Router}. The two handlers are responsible for translating and forwarding queries/messages between different components of the IoT system and the blockchain network. 


We further evaluate the performance of \our{}. 
The experimental results shows that committing and fetching data increases in a quasi linear fraction to the growth of data size. This shows that the overhead incurred by our instantiations can be well tolerated given reasonable data sizes.

%% file: background.tex
\section{Background}

\subsection{Blockchain and Smart Contracts}
Blockchain is an implementation of distributed ledger technology. Every transaction is recorded in the ledger in order of occurrence. In blockchain, a group of transactions is recorded in a block. Blocks are chained by including cryptographic hash value of previous block into the newly created next block. This technique of chaining with linked hash values prevents tampering transactions without being detected.

Smart contracts are computer programs (code) that handle the business logic which was pre-agreed by the network members. They run on the blockchain and provide an interface to interact with the data. This code is available to all the members present on the network (i.e. orderers, peers). Blockchain members add smart contracts to the blockchain in similar way of adding transactions, thus these smart contracts are included in blocks. Transactions that update smart contract states are also recorded in the next block created after the changed had been made. This mechanism makes smart contracts immutable in the same manner for transactions.

Smart contracts are typically enforced by the nodes of the system, therefore is not possible for a single entity to bypass the rules defined within this code, since it would require the agreement of the majority of the participants.
The main advantage of smart contracts is that they can automate an organization's business logic. In turn, the switch to automation cancels the effects of human errors and misunderstandings that may lead to legal disputes. A legal contract or a law might be subject to personal interpretations, but software is deterministic; there is no room for subjective interpretations.

\subsection{Prefix Encryption}
Identity Based Encryption is a cryptographic primitive that solves the problem of authenticity of public keys by replacing the latter with ``identities''. An identity may be an arbitrary string, e.g., an email address like \texttt{jdoe@email.com}. A trusted key-distribution authority provides eligible parties with the secret decryption keys---for example the owner of \texttt{jdoe@email.com} may be granted the corresponding secret decryption key. As a result, one can encrypt data under an arbitrary identity. Assuming a trusted key-distribution authority, none but the party associated to that identity can decrypt the ciphertext and recover the original data.

Prefix encryption~\cite{Boneh:2006:CSI:1272948.1272952} is a special flavor of IBE where decryption keys for a given identity $ID$, allow to decrypt any ciphertext encrypted under identity $ID'$, if $ID$ is a prefix of $ID'$. (Note that any string can be seen as the prefix of itself.). Just like an IBE scheme, a prefix encryption scheme is composed of the following algorithms.

\begin{itemize}

  \item \textbf{$(mpk,msk)\gets Setup(1^\lambda)$.} The setup algorithm takes the security parameter $\lambda$ and outputs a master public key $mpk$ and a master secret key $msk$.

  \item \textbf{$sk_{ID}\gets Extract(msk,ID)$.} The key extraction algorithm takes the master secret key $msk$ and an identity $ID$ and outputs a secret decryption key $sk_{ID}$.

  \item \textbf{$c\gets Encrypt(mpk,ID,m)$.} The encryption algorithm takes the master public key $mpk$ an identity $ID$ and a message $m$; it outputs a ciphertext $c$.

  \item \textbf{$(m,\bot)\gets Decrypt(sk_{ID},c)$.} The decryption algorithm takes as input a secret key $sk_{ID}$ and a ciphertext and outputs either a message $m$ or a special symbol $\bot$.

\end{itemize}

The correctness requirement states that for any message $m$, identity $ID$ and keys $(mpk,msk)\gets Setup(1^\lambda)$, if $sk_{ID}\gets Extract(msk,ID)$, $c\gets Encrypt(mpk,ID',m)$, and  $ID$ is a prefix of $ID'$, $then m\gets Decrypt(sk_{ID},c)$.

The security requirement (IND-ID-CCA) ensures that an adversary cannot gain any information from a ciphertext encrypted under an adversary-chosen identity $ID$, as long as the adversary has access to any decryption key but the ones associated to any identity $ID'$ such that $ID'$ is a prefix of $ID$. 

%% file: system.tex
\section{Secure and Decentralized IoT Data Sharing}
\label{sec:system}

\subsection{Problem Statement}
In this paper, we tackle the design of a secure data-sharing platform for IoT applications. Current IoT platforms place great trust in a single entity (e.g., the IoT broker) who stores all the data, handles all access policies, and takes all access control decisions. 
Few solutions have addressed trust distribution among IoT entities by shifting access control policies to a blockchain back-end. Even so, it still requires every single data owner to perform operations on updating policies, thus making it not scalable to increasingly growth of IoT systems. 

Our overall goal is to decentralize the functionality of a single party in order to guarantee that access policies are correctly managed and evaluated against access requests. We also want to allow auditability of policy updates, as well as access requests and decision. Further, we want to mitigate damages due to data leaks. Finally, we seek for a design where data producers obtain remuneration for sharing their data.

%
%
%
%
%

\subsection{System and Threat Model}

We consider a blockchain comprising of data owners (producers) and data consumers. Data owners primarily store their data on a dedicated remote data storage (e.g., Amazon S3) that is able to harness and process larger amounts of data. Data consumers are interested in accessing data produced by owners. Over blockchains, all members can be either a data consumer, or a data producer, or both.

We assume that data owners control---and rightfully so---are expecting compensation for sharing their data. As such, we assume data to have a ``price'' set by the owner that should be paid by a consumer upon getting access to the data. Both owners and consumers hold unique membership identities over the blockchain network. Identity establishment relies on existing identity manager that often be a part of the blockchain implementation. The identity manager generates and manages identities, keys (public and private), and addresses for all nodes in the network.

In our threat model, we assume data producer and data consumer do not trust each other, however, both of them trust the blockchain network. As by design, blockchain ensures verifiability and immutability, hence its users utilize these features for security. Before sharing data with a consumer, the data producer and the consumer establish a ``payment" transaction given a data offer created previously by the owner. The payment is recorded and verifiable over the blockchain. 

We also assume an active attacker who wants to alter ACLs to get unauthorized access to the data. Note that ACLs is updated according to payment transaction results, thus this attacker needs to control more than a half of voting power of the blockchain network.

%% file: sharing.tex
\section{Our Proposal Framework: \our{}}
\label{sec:sharing}

\subsection{Overview}
We start by giving an overview of \our{}. In \our{}, we store IoT data off-chain but offload to the blockchain the access control functionality---currently handled by a centralized entity such as the IoT broker. 

In particular, a smart contract handles access control policies and evaluates access requests. Data owners push their data to the off-chain storage and advertise it to the smart contract through an ``offer''. The latter may define the price to be paid in order to gain access to the data. Similarly, access requests from consumers are issued to the smart contract that evaluates the request against the policy and makes an access decision. The smart contract also bookkeeps trade information between owners and consumers via IOU\footnote{IOU is abbreviated from the phrase "I owe you". It usually specifies a debtor, the amount owed, and might be also creditor} accounts. As such, our platform creates a ``data marketplace'' where owners sell and consumers buy data.

As we handle access control in the blockchain, we ensure that access policies are correctly managed and access requests are duly evaluated. We also ensure remuneration of data owners and auditability of all operations.

Since data is actually stored off-chain, the storage provider is also a blockchain node executing the smart contract. Once an access decision is made, the storage provider acts accordingly---allowing or denying access to the data---thereby performing as the policy enforcement point.

Up to this point, our solution overview assumes a trusted storage provider. Nevertheless, a malicious storage provider may abuse its functionality as policy enforcement point and, e.g, share data with unauthorized parties. We address this issue in our extended design where 1) data is encrypted before uploading it to the storage provider, and 2) key distribution to authorized parties is handled by a cohort of key authorities. As a result, the storage provider only handle ciphertexts and unauthorized access requires compromise of all the parties acting as key authorities. Given the application scenario where data (and sensors) is usually organized in a hierarchical fashion, we identify prefix encryption as a suitable encryption scheme that allows fine-grained access control while minimizing key requests to the (distributed) authority.

\subsection{Data Marketplace}
The data marketplace in \our{} allows data producers and consumers to trade data through the blockchain. Here, we design basic trading functions over the marketplace: \textbf{verifyMetaData()}, \textbf{addMetaData()}, \textbf{createOffer()} and \textbf{acceptOffer()}. Data structures of the various entities are shown in Listing~\ref{datastructure}. In the following we will present details of data marketplace functions.


\begin{minipage}{\linewidth}
\begin{lstlisting}[label=datastructure,caption=Data Structure]
Struct User           Struct MetaData
    ID: string             ID: string
    Pk: []byte             fileID: []byte
                           owner: []byte
                           URI: []string
                           ACL:[][]byte
Struct Offer           Struct IOU
    ID: string             ID: string
    mdata: *MetaData        User1: string
    value: float           User2: string
    state: bool            value: float
\end{lstlisting}
\end{minipage}

\subsubsection{Data structure}

Each \texttt{MetaData} item has unique fileID, owner information, storage location and white-list type ACL (Listing~\ref{datastructure}). Initially, the ACL only contains the owner. Owners can share their data by creating an offer that sets the price to get access to the data. Data offers contain a reference to the data being sold and these offers are available for trading only when their state is ``true'' (active). Available offers are visible on the blockchain so that consumers can query and find the data that fits their interests. To handle payments, an IOU account is created for each pair of seller (who created data offer) and buyer (who accepts offer and pays for getting access). IOU accounts are updated to reflect results of transactions between the trading parties. IOUs can be settled off-chain using standard payment methods. How these IOUs can be used and converted to other currencies depends on specific implementation of the blockchain and it is out of our scope. 

\begin{minipage}{\linewidth}
\begin{lstlisting}[label=verifydata,caption=Data verification function]
Function verifyData(user, mdata, cloud)
    foreach file in mdata do
        if !cloud.exists(file.URI) or
           cloud.getID(file.URI) != file.ID or
           cloud.getACL(file.URI) != file.ACL or
           cloud.getOwner(file.URI) != file.owner
        then
            return false
        end if
    end for
    return true
end function
\end{lstlisting}
\end{minipage}

\subsubsection{Verify meta data}
Listing~\ref{verifydata} describes the function to verify meta data. It checks the match of metadata between a given data and respective details of the same data stored at the cloud. This verification does not check quality of data content.

\subsubsection{Add meta data to the blockchain}
The blockchain network allows its nodes to add meta data that they want to share. It stores this data in a key-value store database. To add meta data, it first verifies the meta data. Next it adds the data owner to the ACL (see Listing~\ref{adddata}). 

\begin{minipage}{\linewidth}
\begin{lstlisting}[label=adddata,caption=Data adding function]
Function addData(user, mdata, cloud)
    if !verifyData(user, mdata, cloud) then
        return error
    end if
    if map.Contains(mdata.fileID) then
        return error
    end if
    data.ACL = {mdata.owner}
    key = "data" || mdata.fileID
    map.Set(key, mdata)
    return success
end function
\end{lstlisting}
\end{minipage}

\subsubsection{Create data offer}
To share data with an interested consumer, the data owner needs to create an offer for that data and let the network advertise this offer to potential consumers. The value field represents the price that the data can be traded for. Listing~\ref{createoffer} contains steps in a smart contract to allow a node adding an offer to the blockchain. Before an offer for a data is added to the blockchain, the function data verification is called to check for data validity. Once the verification passed, the offer is added to the blockchain and its state is set ``active'' making the offer available for trading. Data offers can be revoked (or canceled) by setting their state values to false. To put focus on its functionality, we consider simple offers with few basic data fields, however, in practice, the offer might have more fields such as the validity period and accomodate composite sharing conditions.

\begin{minipage}{\linewidth}
\begin{lstlisting}[label=createoffer,caption=Offer creating function]
Function createOffer(user, offer, cloud)
    if !verifyData(user, offer.mdata, cloud) then
        return error
    end if
    offer.state = true
    key = "offer"||offer.mdata.fileID
    map.Set(key, offer)
    return success
end function
\end{lstlisting}
\end{minipage}

\subsubsection{Accept data offer}
Data consumers can browse all available offers by querying the blockchain. Once a data item of interest is found, the consumer can accept the offer, make the payment and download the data. Listing~\ref{acceptoffer} describes the corresponding function. When the consumer accepts the data offer by the owner, the owner-consumer IOU account is updated.



\begin{minipage}{\linewidth}
\begin{lstlisting}[label=acceptoffer,caption=Offer accepting function]
Function acceptOffer(user, offer, cloud)
   if !offer.state then
     return error
   end if
   if !verifyData(offer.mdata.owner, offer.mdata, 
     cloud) then
     return error
   end if
   offer.state = false
   key = offer.ID
   map.Set(key, offer)
   if user.ID > offer.mdata.owner then
     key = "IOU"||hash(user.ID, offer.mdata.owner)
     cvalue = getCurrentIOTValue(key)
     value = cvalue + offer.value
   else
     key = "IOU"||hash(user.ID, offer.mdata.owner)
     cvalue = getCurrentIOTValue(key)
     value = cvalue - offer.value
   end if
   map.Set(key, append(offer.mdata.ACL, user.ID))
   return success
end function
\end{lstlisting}
\end{minipage}

\subsection{Data Sharing Schemes}
\label{sec:schemes}

Our description thus far covers data trading operations. In the following we describe how data is actually shared, i.e., how consumers get access to data they have requested. We design two schemes based on Access Control Lists (ACLs) and on prefix encryption, respectively

\subsubsection{ACLs based scheme}
\label{sec:proposal:acl}

\begin{figure}[t]
    \centering
    \includegraphics[width=0.5\textwidth]{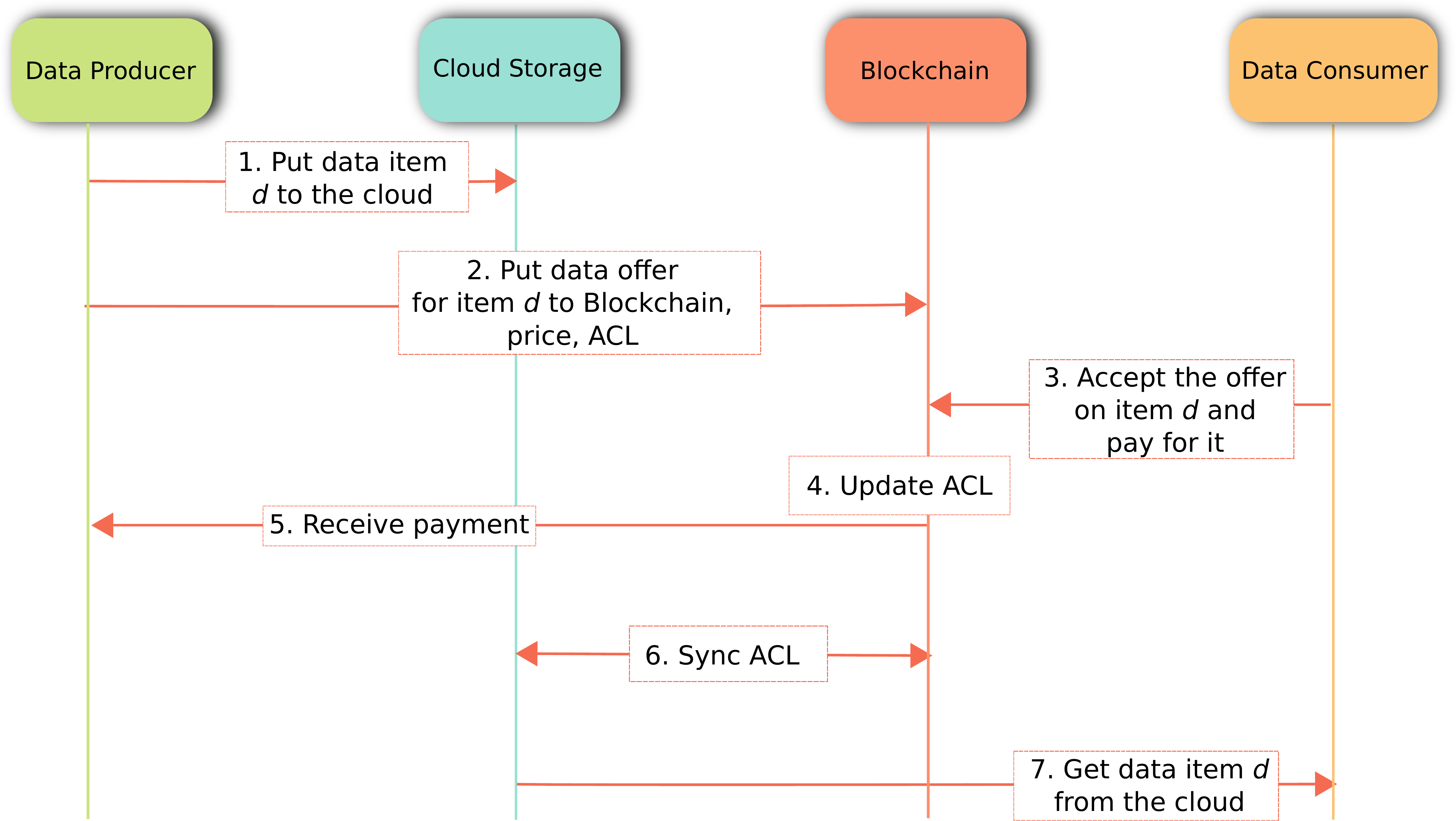}
    \caption{Data sharing scheme based on ACL.}
    \label{fig:acl}
\end{figure}

An Access Control List (ACL) is a list of permissions to manage who can access a data resource. A data owner stores its data in the clear at the cloud storage that holds the latest version of the ACL---since the storage provider is also a node of the blockchain---and enforces access control decisions. Fig.\ref{fig:acl} depicts the sharing steps where producer and consumer advertise and accept data offers. The data offer is described in a smart contract and the corresponding ACL is updated by adding the identity of the consumer once its access request has been accepted. Note that ACL updates are done over the blockchain (by executing the martketplace smart contract) and do not require operations from data owners.

Enforcement via ACLs provides a straightforward means to regulate access to data. Nevertheless, this scheme assumes a trusted storage provider that does not abuse its role as a policy enforcement point to, e.g., leak data to unauthorized parties.

\subsubsection{Prefix encryption based scheme}
\label{sec:proposal:ibe}

\begin{figure}[t]
    \centering
    \includegraphics[width=0.5\textwidth]{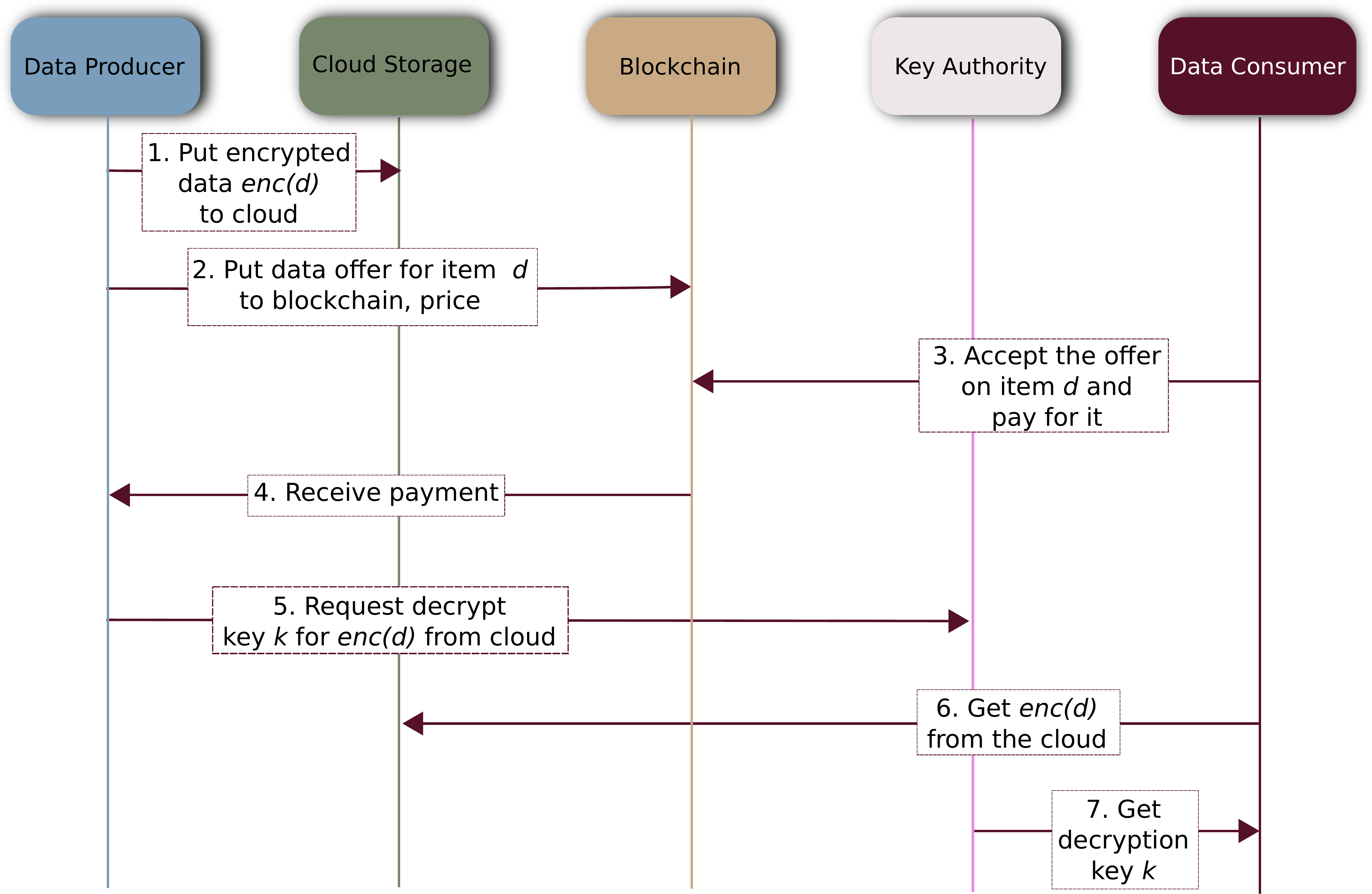}
    \caption{Data sharing scheme based on prefix encryption.}
    \label{fig:ibe}
\end{figure}
We now introduce a scheme where access control is cryptographically enforced so that we can refine the trust assumption on the storage provider. In particular we assume data is encrypted by owners before it is uploaded to the storage provider and we use the blockchain-managed ACL to reflect parties authorized to access the corresponding decryption keys. Fig.\ref{fig:ibe} provides an overview of this design. Here the cloud is detached from the blockchain network.

We use prefix encryption as it is particularly suited for IoT scenarios where a multitude of devices are organized in a hierarchy. 
For example, Alice may organize her IoT devices in an hierarchical namespace \texttt{alice/}, including \texttt{alice/house/} for devices installed at her house and \texttt{alice/car/} for devices installed in her car. Alice can thus create her own master key-pair (i.e., by running $Setup(1^\lambda)$) and load the master public key on her devices. Each device produces data encrypted under its own ``prefix''. For example, the smart thermostat may encrypt its measurement $m$ via $Encrypt(mpk,\texttt{alice/home/thermostat},m)$ and upload the ciphertext to the platform. Envelope encryption may be used to encrypt bulk data.

Distribution of secret keys to eligible parties can happen in a number of ways. Alice may setup her own service or, alternatively, delegate this role to either a single or a distributed authority. The party taking up such a role must hold the master secret key $msk$ created by Alice at $Setup$ time\footnote{If the authority is distributed, each peer receives a share of the master secret key.} and must synchronize with the blockchain to get the up-to-date ACL for a given prefix and distribute decryption keys accordingly. The key authority receives a request to access a given prefix, e.g., \texttt{alice/home/thermostat}) and uses the ACL to decide whether to grant or deny the request. If the request is granted, the authority runs $Extract$ to compute the secret decryption key and securely transfer that key to the requestor. Note that requests carry the public key of the requestor so that the authority can securely transfer keys to the intended party.

Once a party obtains a decryption key associated to a given prefix it can ask for data produced under that prefix to the storage provider. The latter does not manage cleartext data as devices only upload encrypted data. Therefore, the storage provider does not carry out any policy enforcement but simply forwards the ciphertexts to the requestor. Finally the requestor holding the decryption key runs the $Decrypt$ algorithm and recovers the cleartext data. Note a requestor may use the decryption key for a prefix (e.g., \texttt{alice/home/}) to decrypt ciphertexts produced by any device under that prefix (e.g., \texttt{alice/home/thermostat} or \texttt{alice/home/doorlock}).

\subsection{Security Analysis}


In the two data sharing schemes (ACLs based and prefix encryption based), a rational cloud policy decision point cannot influence the
consensus in the blockchain (i.e., the data owner's decision). Recall that the owners vote on access control decisions by issuing appropriate blockchain transactions. Such transactions are confirmed in the blockchain by the validators/miners.
As required for the security of the underlying blockchain, we
assume the standard safety conditions particular to the underlying
blockchain technology. For instance, in Proof-of-Work (PoW) based
blockchains (e.g., Bitcoin and Ethereum), we assume that the
adversary cannot control the majority of the computing power in
the network. Note that the access
control decisions made over the blockchain and enforced the cloud PEP according to the user
contract which was previously agreed upon by all owners. These
decisions are publicly verifiable and the cloud provider can be held
accountable for any diverging decisions.

Further, the blockchain ensures that all operations (i.e., data offers, offer accepts, ACL updates and changes to IOU accounts) are auditable. The basic schemes of Section~\ref{sec:schemes} ensure that access to data is correctly enforced---only parties with an identity in the blockchain and that have paid the price to access the data---assuming a trusted storage provider. The scheme that uses prefix encryption can tolerate a malicious storage provider and requires a trusted key distribution authority. Nevertheless, the authority can be distributed across several peers so that no data is leaked unless \emph{all} peers are malicious. We note that trivial DoS attacks---for example, the storage provider may simply deny access to (encrypted) data---are out of our scope.

\subsection{Integration: \our, FIWARE and Hyperledger Fabric}
\label{sec:proposal:integration}

\begin{figure}[t]
    \centering
    \includegraphics[width=0.3\textwidth]{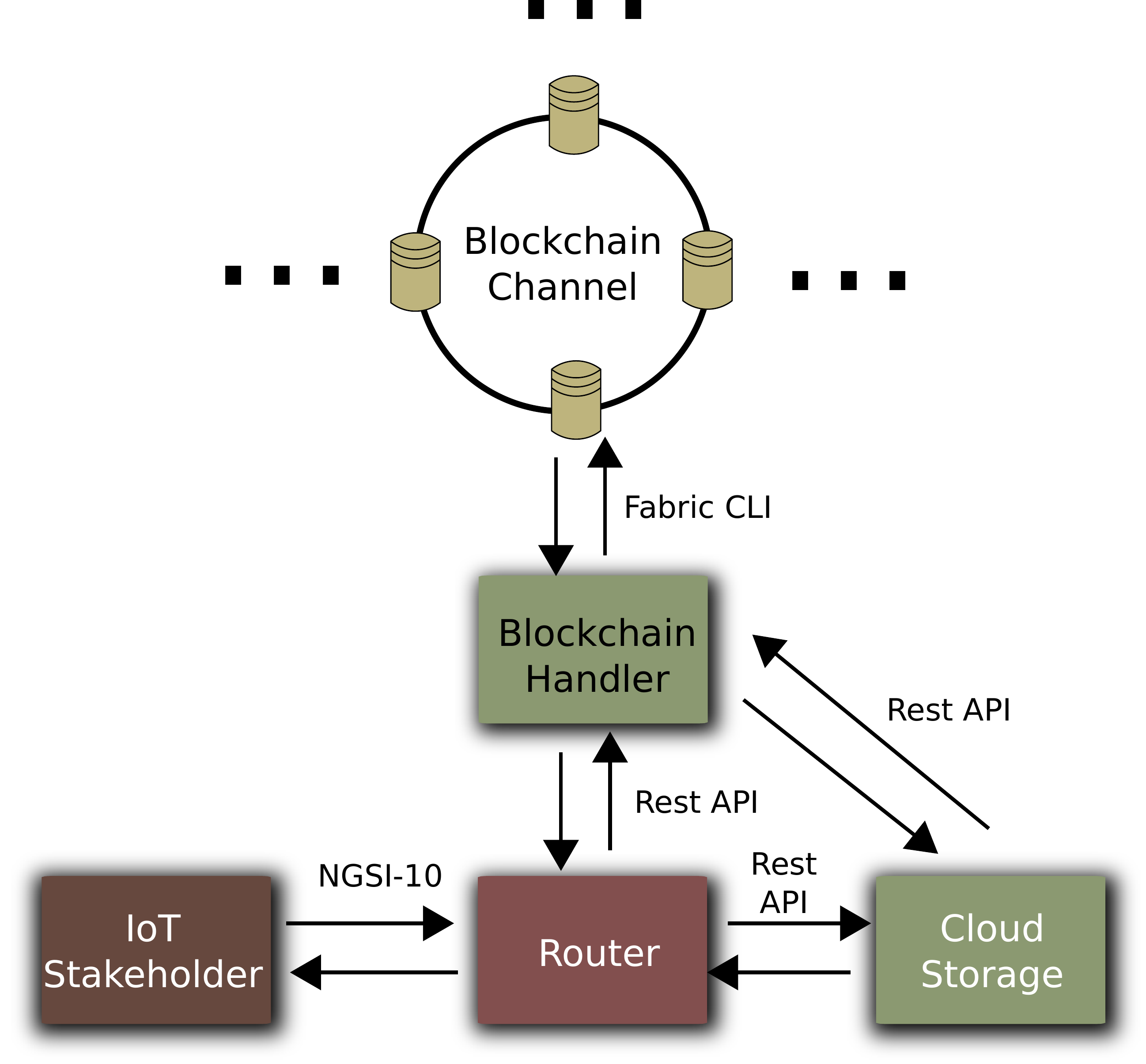}
    \caption{Integration architecture.}
    \label{fig:integration}
\end{figure}


FIWARE, created in Europe from the Future Internet Public Private Partnership (PPP), is an IoT open platform to provide common APIs that ease the development of smart applications. It allows to connect vendors, developers, and data providers to collect, process and analyze context data at large scale.
The core of open FIWARE are generic enablers (GEs). A FIWARE generic enabler is a set of general-purpose platform functions available to application developers through APIs.
FIWARE is a cloud-based platform and consists of two main parts: IoT Edge and IoT back-end. Operating on top of IoT back-end is Data Context Broker component which is connected to applications. IoT Edge supports IoT gateway and IoT NGSI~\footnote{http://aeronbroker.github.io/Aeron/} gateway for NGSI-capable devices.

FIWARE includes following major entities: \textit{cloud} which refers to federation infrastructures to deploy GEs; \textit{data} which refers to context management platform and the integration of data; \textit{IoT} which refers to multiple protocols to plug and play IoT devices; \textit{apps} which to publish and visualize data; \textit{Web UI} is a support of UIs to represent context information; \textit{I2ND} which provides networking capabilities; and \textit{security} which provides built-in access/identity/privacy management.

Hyperledger Fabric (simply \emph{Fabric}) is an open-source blockchain platform~\cite{Androulaki:2018:HFD:3190508.3190538} managed by the Linux Foundation~\footnote{\url{https://www.hyperledger.org}}. Fabric has widely range of use in prototypes, proof-of-concepts and industrial production. Use cases of Fabric include various areas such as supply chain management, contract management, data provenance, identity management. 

Fabric uses hybrid replication design which incorporates primary-backup (passive) replication and active replication. Primary-backup replication in Fabric means every transaction is executed only by a subset of peers based on endorsement policies. Fabric adopts active replication that transactions are written to the ledger once reaching consensus of total order. This hybrid design makes Fabric a scalable permissioned blockchain.

\our{} includes five entities: \texttt{IoT Domain} which contains \texttt{IoT Stakeholder}, \texttt{Cloud Storage}, \texttt{Blockchain} and \texttt{Key Authority} (optional). We consider \texttt{IoT Domain} as a virtual concept for a group of \texttt{IoT stakeholders} that share common settings (e.g. located in a same building, produce same type of context data). The IoT Domain acts as a node that can be both data producer and data consumer. It can be referred as IoT middleware layer component which provides capabilities to connect to further systems and applications that low-level IoT hardware-based components lack of. Each IoT Domain maintains a router to forward communications from IoT stakeholders to further applications. Our extension is done by adding two new handlers that we name \texttt{Blockchain Handler} and \texttt{Router}. These additions come from the fact that components are implemented in different languages following various standards and thus these components cannot communicate directly to each other. 
The hetogeneousity becomes even larger when we consider various cloud providers which have their own web APIs. Fig.\ref{fig:integration} gives an overview of components and their interactions in \our{}.

When IoT stakeholders connect to their IoT Domain router, the router forwards the data stream to a right handler. That is, when the message belongs to IoT data streams (files) types, it is directed to the cloud storage. When the message belongs to query types for retrieving data from the IoT to perform a particular physical action by IoT stakeholders, it is forwarded to the blockchain handler for verifying access permissions. Here IoT domains join the blockchain network and query ACLs for the requested data item. As we have introduced two sharing schemes, the design of either having an ACL or a key authority is adjustable in this integration. Similarly, whether or not the cloud storage entity joins the blockchain network (data marketplace) is also adjustable without any required changes on the architecture of existing IoT platforms. We argue that \our{} can be integrated to other IoT platforms with proper adjustments.

Communications (sending/receiving messages) between components are mediated by handlers. An IoT Domain acts as a node in the blockchain. The blockchain has one node that runs the web RESTful API and translates standard HTTP (e.g. GET, POST) requests to queries to the smart contracts. The replies from the contract are added to the HTTP response body. Each node implements a router that interprets FIWARE NGSI-10 requests and translates them into the commands that the Web API the blockchain handler expects. In addition, this service is also responsible for the communication with the cloud via its web APIs. The blockchain handler is responsible for all communications to the blockchain network via provided APIs. We adopt the implementation of Hyperledger Fabric and use its blockchain APIs e.g., Fabric CLI for connection between blockchain handler and blockchain network.



%


%% file: evaluation.tex
\section{Implementation \& Evaluation}
\label{sec:eval}

\subsection{Implementation Setup}

In our implementation, the majority of the components were developed using Golang as the programming language. We implemented a smart contract to run the data sharing functions described in Section~\ref{sec:sharing}. And we deployed this contract in a Fabric network using Kafka as the consensus algorithm. 

In our experiments, we evaluate the efficiency \our{} focusing on two data-sharing schemes. The experimental setup consists of a blockchain network built from HyperLedger Fabric modules. This network has one ordering service and five connecting peers. One of the peers runs the web API for the entire blockchain network. Regarding the two sharing schemes, we identify that the data size and the depth of the encryption key (for the prefix encryption based scheme) will affect the system's performance. Our measuring metric is the time it takes to commit the data to the network (for data producers) and the time it takes to fetch the data from the cloud (for consumers). The data is being sent from a machine connected to the same network as the peer, and the cloud service provider is mimicked by a key/value store database hosted by the same machine as the peer.

For the key depth used in the prefix encryption based scheme, we tested different depth levels from 1 to 6 and there are no differences in performance besides experimental noise. Thus we conclude that the performance of the prefix encryption based scheme does not change with the key depths.

For the data size, we performed the aforementioned test using data arrays that are of size 1, 5, 10 and 20 MB. For each, we generated a random array of bytes and ran the queries to submit and fetch the data from/to the system.

\subsection{Evaluation Results}
\begin{figure}[t]
    \centering
    \includegraphics[width=0.5\textwidth]{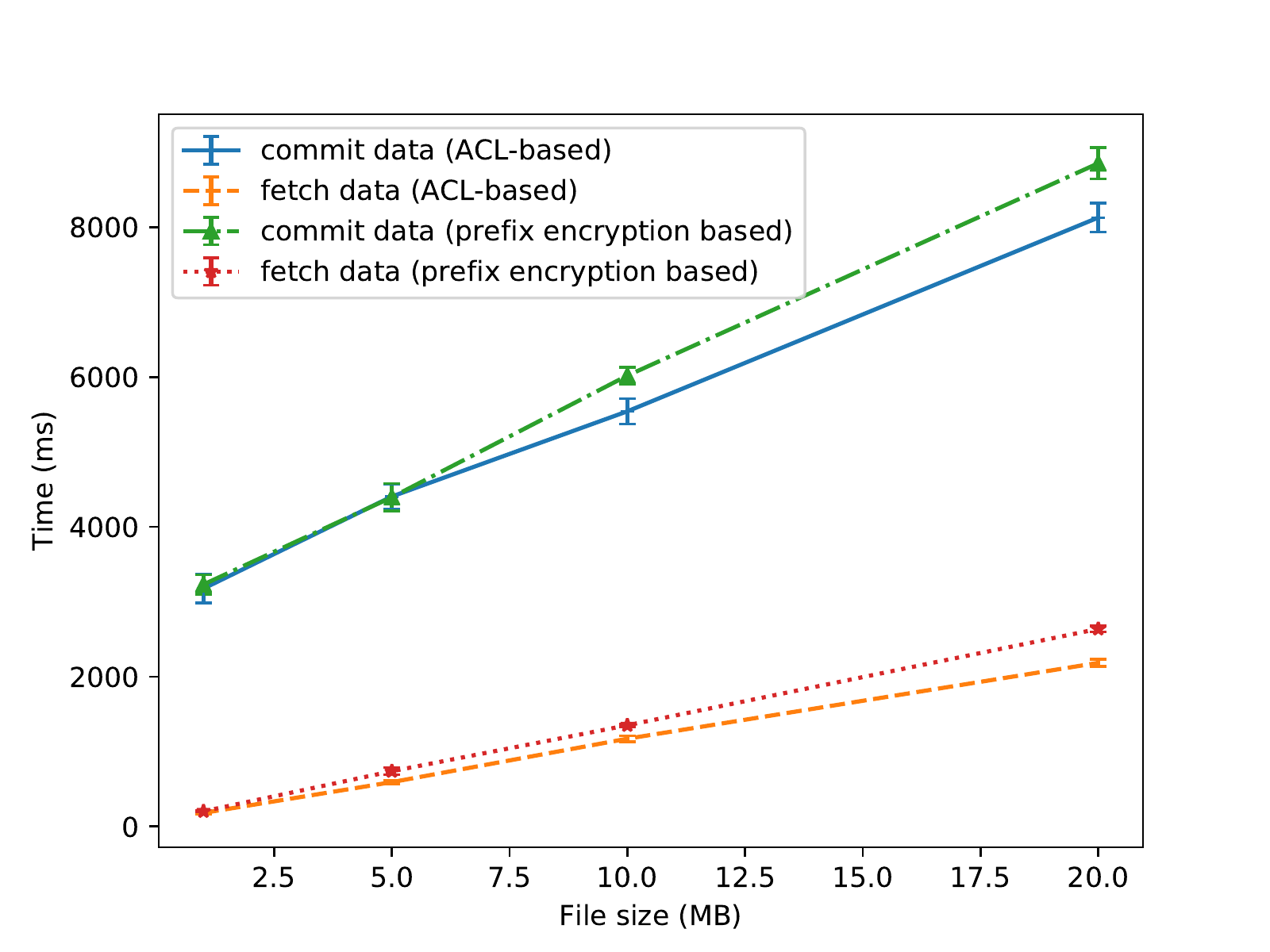}
    \caption{Performance results with variance of data sizes.}
    \label{fig:result}
\end{figure}

The experimental results (Fig.\ref{fig:result}) show that committing data to the platform takes longer than fetching it. This can be explained by the added complexity (number of smart contract calls) of one action versus the other. In a real world deployment we speculate that the number of fetching actions  will be higher than that of commit actions, meaning that more users will act as data consumers than as data producers. Therefore mitigating the throughput gap between these two actions.

We can also see that, as expected, the time to perform any of the evaluated actions increases in a quasi linear fraction with the increase in the size of the data. This can clearly be attributed to the time it takes for the data to travel through the network.
Furthermore, whilst the penalty from using encryption is almost imperceptible for smaller data sizes which becomes more impactful as the data size increases. We argue that this is a worthwhile trade off given the benefits brought by prefix encryption.

%% file: related.tex
\section{Related Work}
Researchers have proposed the use of blockchain for various IoT domains such as the Internet of healthcare things~\cite{Esposito18}, the Internet of vehicles~\cite{Qiu:2018:BDS:3272036.3272040, Lei17}, the Internet of energy~\cite{Li18, Liang18}, and crowdsensing~\cite{2018arXiv180804056C}. In~\cite{2018arXiv180606185P} Wang et al., designed and prototyped an edge IoT framework ``EdgeChain'' based on permissioned blockchain and currency mechanism to manage how much resource IoT devices can get from edge servers. 
IoT industry players also have adopted blockchain and smart contracts to build numerous applications (e.g. \textit{Slock.it}~\footnote{\url{https://slock.it/}} -- a German-based company that builds an application for renting, sharing and selling without involvement of any trusted third party; \textit{Ujo}~\footnote{\url{https://www.ujomusic.com/}} uses blockchain to handle ownership rights such that music owners get payment once their music is used for commercial purpose).

Following the trend of leveraging blockchain for IoT, a number of schemes for secure
sharing of data over blockchains have been explored by the researcher community~\cite{Shafagh:2017:TBA:3140649.3140656, cryptoeprint:2018:209, Christidis16}. In~\cite{cryptoeprint:2018:209} Kokoris-Kogias et al., proposed CALYPSO for auditable sharing of private data where data is stored onchain and collective authorities formed over the blockchain are responsible for enforcing access control policies. This design is not suitable for dealing with huge amount of IoT data generated by large numbers of IoT devices in real-world systems. Shafagh et al.,~\cite{Shafagh:2017:TBA:3140649.3140656} designed a system for sharing time-series IoT data where data owners have to issue transactions for setting policies each time the data is shared with another party and only the owner can change that policy later. Similarly, Laurent et al.,~\cite{Laurent2018AnAC} proposed a system where blockchain is used to handle transactions between parties before granting permissions, however how such transactions are done is not addressed, and only owners can change policies. Our design allows ACL updates or decryption key distributions to be autonomously done over the blockchain back-end without any data owner's interventions. Owners only specify for data offers, trading and granting are handled by the blockchain.

Data monetisation has been an active research topic~\cite{Suliman2019,Shrestha18,Mylrea2017}. This trend is supported by the growth of machine learning and smart technologies. It allows IoT devices and systems to generate revenue from the data they produce. In~\cite{Suliman2019} Suliman et al., implemented Ethereum smart contracts for monetizing IoT data. Shrestha and Vassileva~\cite{Shrestha18} proposed a framework for research data sharing that provides incentives to data owners. Mylrea and Gourisetti~\cite{Mylrea2017} introduced applications of blockchain and smart contract for Energy-IoT data, to enable trading energy data in smart grids.
Even though these works targeted monetisation in different IoT domains, they all only focused on building mechanisms for rewarding data owners. In \our{}, in addition to providing ``by design'' data owners rewards, we also integrate such  data monetisation as a part of data sharing schemes where access control policies and key distribution are decided based on how trading transactions have been completed.

%% file: conclusion.tex
\section{Concluding Remarks}
We have presented \our{}, the first secure and decentralized IoT data-sharing framework over blockchains that provides auditable access control policy updates and data owner remuneration. \our{} achieves its goal by introducing three main components in its design: a data marketplace and two sharing schemes based on ACLs and prefix encryption. The data marketplace allows a large number of data owners and consumers to trade on IoT data. The two sharing schemes provide different ways to grant access policies, one is based on ACLs and the other is based on distribution of prefix decryption keys. We have implemented \our{} using open source platforms FIWARE and Hyperledger Fabric and shown that open IoT platforms can enhance their security by integrating \our{} as an extension. Initial performance evaluation results show a moderate overhead.

In term of future work, we plan to apply \our{} to allow for
validating global policies in the network. Namely,
existing IoT platforms feature a search engine that connects to multiple IoT
domains to index and crawl IoT data. These multiple
domains do not trust each other and yet need to agree on a set of
policies for sharing data indexed by the search engine. We believe that \our{} may find immediate applicability in this context.

%% file: ack.tex
\section*{Acknowledgment}
This paper has received funding from the European Union’s Horizon 2020 research and innovation programme under grant agreement No 779852.

%% file: main-arxiv.bbl
\begin{thebibliography}{10}

\bibitem{Androulaki:2018:HFD:3190508.3190538}
E.~Androulaki, A.~Barger, V.~Bortnikov, C.~Cachin, K.~Christidis, A.~De~Caro,
  D.~Enyeart, C.~Ferris, G.~Laventman, Y.~Manevich, S.~Muralidharan, C.~Murthy,
  B.~Nguyen, M.~Sethi, G.~Singh, K.~Smith, A.~Sorniotti, C.~Stathakopoulou,
  M.~Vukoli\'{c}, S.~W. Cocco, and J.~Yellick.
\newblock Hyperledger fabric: A distributed operating system for permissioned
  blockchains.
\newblock In {\em Proceedings of the Thirteenth EuroSys Conference}, EuroSys
  '18, pages 30:1--30:15, New York, NY, USA, 2018. ACM.

\bibitem{Boneh:2005:HIB:2154598.2154634}
D.~Boneh, X.~Boyen, and E.-J. Goh.
\newblock Hierarchical identity based encryption with constant size ciphertext.
\newblock In {\em Proceedings of the 24th Annual International Conference on
  Theory and Applications of Cryptographic Techniques}, EUROCRYPT'05, pages
  440--456, Berlin, Heidelberg, 2005. Springer-Verlag.

\bibitem{Boneh:2006:CSI:1272948.1272952}
D.~Boneh, R.~Canetti, S.~Halevi, and J.~Katz.
\newblock Chosen-ciphertext security from identity-based encryption.
\newblock {\em SIAM J. Comput.}, 36(5):1301--1328, Dec. 2006.

\bibitem{2018arXiv180804056C}
D.~{Chatzopoulos}, S.~{Gujar}, B.~{Faltings}, and P.~{Hui}.
\newblock {Privacy Preserving and Cost Optimal Mobile Crowdsensing using Smart
  Contracts on Blockchain}.
\newblock {\em ArXiv e-prints}, Aug. 2018.

\bibitem{Christidis16}
K.~Christidis and M.~Devetsikiotis.
\newblock Blockchains and smart contracts for the internet of things.
\newblock {\em IEEE Access}, 4:2292--2303, 2016.

\bibitem{Esposito18}
C.~Esposito, A.~D. Santis, G.~Tortora, H.~Chang, and K.~R. Choo.
\newblock Blockchain: A panacea for healthcare cloud-based data security and
  privacy?
\newblock {\em IEEE Cloud Computing}, 5(1):31--37, Jan./Feb. 2018.

\bibitem{Hu2018}
S.~{Hu}, C.~{Cai}, Q.~{Wang}, C.~{Wang}, X.~{Luo}, and K.~{Ren}.
\newblock Searching an encrypted cloud meets blockchain: A decentralized,
  reliable and fair realization.
\newblock In {\em IEEE INFOCOM 2018 - IEEE Conference on Computer
  Communications}, pages 792--800, April 2018.

\bibitem{Kaaniche2017}
N.~{Kaaniche} and M.~{Laurent}.
\newblock A blockchain-based data usage auditing architecture with enhanced
  privacy and availability.
\newblock In {\em 2017 IEEE 16th International Symposium on Network Computing
  and Applications (NCA)}, pages 1--5, Oct 2017.

\bibitem{cryptoeprint:2018:209}
E.~Kokoris-Kogias, E.~C. Alp, S.~D. Siby, N.~Gailly, L.~Gasser, P.~Jovanovic,
  E.~Syta, and B.~Ford.
\newblock Calypso: Auditable sharing of private data over blockchains.
\newblock Cryptology ePrint Archive, Report 2018/209, 2018.
\newblock \url{https://eprint.iacr.org/2018/209}.

\bibitem{Laurent2018AnAC}
M.~Laurent, N.~Kaaniche, C.-Y. Le, and M.~V. Plaetse.
\newblock An access control scheme based on blockchain technology.
\newblock 2018.

\bibitem{Lei17}
A.~Lei, H.~Cruickshank, Y.~Cao, C.~P.~Anyigor~Ogah, P.~Asuquo, and Z.~Sun.
\newblock Blockchain-based dynamic key management for heterogeneous intelligent
  transportation systems.
\newblock PP, 08 2017.

\bibitem{Li18}
Z.~Li, J.~Kang, R.~Yu, D.~Ye, Q.~Deng, and Y.~Zhang.
\newblock Consortium blockchain for secure energy trading in industrial
  internet of things.
\newblock {\em IEEE Transactions on Industrial Informatics}, 14(8):3690--3700,
  Aug 2018.

\bibitem{Liang18}
G.~Liang, S.~R. Weller, F.~Luo, J.~Zhao, and Z.~Y. Dong.
\newblock Distributed blockchain-based data protection framework for modern
  power systems against cyber attacks.
\newblock {\em IEEE Transactions on Smart Grid}, pages 1--1, 2018.

\bibitem{Mylrea2017}
M.~{Mylrea} and S.~N.~G. {Gourisetti}.
\newblock Blockchain for smart grid resilience: Exchanging distributed energy
  at speed, scale and security.
\newblock In {\em 2017 Resilience Week (RWS)}, pages 18--23, Sep. 2017.

\bibitem{2018arXiv180606185P}
J.~{Pan}, J.~{Wang}, A.~{Hester}, I.~{Alqerm}, Y.~{Liu}, and Y.~{Zhao}.
\newblock {EdgeChain: An Edge-IoT Framework and Prototype Based on Blockchain
  and Smart Contracts}.
\newblock {\em ArXiv e-prints}, June 2018.

\bibitem{Qiu:2018:BDS:3272036.3272040}
C.~Qiu, F.~R. Yu, F.~Xu, H.~Yao, and C.~Zhao.
\newblock Blockchain-based distributed software-defined vehicular networks via
  deep q-learning.
\newblock In {\em Proceedings of the 8th ACM Symposium on Design and Analysis
  of Intelligent Vehicular Networks and Applications}, DIVANet'18, pages 8--14,
  New York, NY, USA, 2018. ACM.

\bibitem{Shafagh:2017:TBA:3140649.3140656}
H.~Shafagh, L.~Burkhalter, A.~Hithnawi, and S.~Duquennoy.
\newblock Towards blockchain-based auditable storage and sharing of iot data.
\newblock In {\em Proceedings of the 2017 on Cloud Computing Security
  Workshop}, CCSW '17, pages 45--50, New York, NY, USA, 2017. ACM.

\bibitem{Shrestha18}
A.~Shrestha and J.~Vassileva.
\newblock {\em Blockchain-Based Research Data Sharing Framework for
  Incentivizing the Data Owners}, pages 259--266.
\newblock 06 2018.

\bibitem{Suliman2019}
A.~Suliman, Z.~Husain, M.~Abououf, M.~Alblooshi, and K.~Salah.
\newblock Monetization of iot data using smart contracts.
\newblock {\em IET Networks}, 8:32--37(5), January 2019.

\bibitem{Yakubov}
A.~{Yakubov}, W.~M. {Shbair}, A.~{Wallbom}, D.~{Sanda}, and R.~{State}.
\newblock A blockchain-based pki management framework.
\newblock In {\em NOMS 2018 - 2018 IEEE/IFIP Network Operations and Management
  Symposium}, pages 1--6, April 2018.

\end{thebibliography}
